\documentclass[reprint,prb,leqn,showkeys,showpacs]{revtex4-1}
\usepackage{amsmath,amsfonts,amssymb,mathrsfs,graphicx}
\usepackage[colorlinks,linkcolor=blue,citecolor=blue,urlcolor=blue]{hyperref}
\usepackage[usenames,dvipsnames]{color}

\renewcommand\j{\psi}

\newcommand\J{\Psi}

\newcommand\scr{Schr\"odinger }

\def\ket#1{\lvert#1\rangle}

\def\bra#1{\langle #1 \rvert}

\def\amp#1#2{\langle #1 \lvert #2 \rangle}

\newcommand\ra{\rightarrow}

\newcommand{\lan}{\langle}
\newcommand{\ran}{\rangle}

\begin{document}

\title{Do Bloch waves interfere with one another ?}

\author{Vivek M. Vyas}
\affiliation{Indian Institute of Information Technology Vadodara, Government Engineering College, Sector 28, Gandhinagar 382028, India\\Email: vivek.vyas@iiitvadodara.ac.in}

\begin{abstract}
Here we show that two Bloch states, which are energy eigenstates of a quantum periodic potential problem, with different wavevectors can not be linearly superposed to display quantum interference of any kind that captures the relative phase between them. This is due to the existence of a superselection rule in these systems, whose origin lies in the discrete translation symmetry. A topological reason leading to such a superselection is found. A temporal analogue of this superselection rule in periodically driven quantum systems is also uncovered, which forbids the coherent superposition of any two quasi-periodic Floquet states with different quasienergies.
\end{abstract}
\date{\today}
\maketitle

\section{Introduction}

The phenomenon of wave interference has been a subject of intense research since its discovery by Young through the celebrated double slit experiment.\cite{born2013}  For several decades physicists struggled to reconcile the interference phenomena displayed by light, which is also depicted fairly accurately by geometrical optics in many scenarios. But it was only after the advent of quantum mechanics that the interference phenomenon attained its elevated status in physics, while it became an indispensable part of our current understanding of the quantum world.\cite{feynmanbook,diracbook,agarwal2013} 

It is well known that the origin of interference phenomenon in the context of classical light propagation is the fact that the wave equation is a linear partial differential equation. As a result given two propagating solutions of this equation, a linear superposition of the two is also a solution, which results in interference phenomena. The property of linearity is also central to quantum mechanics, since the time evolution of a (closed) quantum system is governed by \scr equation, which is also a linear partial differential equation. As argued elegantly by Feynman, Dirac and others, the interference displayed by quantum systems owing to linearity, is solely of quantum origin, in the sense that the same system operating under classical physics can not display such an effect.\cite{diracbook,feynmanbook,ecgbook} As a result the presence of interference effects in a system is used as a measure of quantum nature of the system.\cite{agarwal2013,gerry2005}

The state space of a quantum system consists of a linear vector space, which allows for arbitrary linear superpositions of the constituent vectors, each of which is a possible quantum state of the system. This may lead one to think that there is an infinite variety of linear superpositions that one can be realised in quantum mechanics, in principle, and hence one can have an infinite variety of interference phenomena which are possible. This belief was found to be factually incorrect in a classic paper by Wick, Wightman and Wigner, wherein it was shown that a linear superposition of any two quantum states does not necessarily leads to an interference phenomenon.\cite{wick1952} It was shown that in a certain class of quantum systems, the Hilbert space of the system gets divided into disjoint sectors, called the \emph{superselection sectors}. The interference phenomena is only displayed if the linearly superposed state is created out of any two states both belonging to same superselection sector. If one is able to create a linear superposition of states belonging to different sectors, counter intuitively, it was shown that no interference effect is displayed by such a state. This lack of coherent superposition of quantum states implies that there is a restriction on the occurrence of quantum interference, which came to be known as the \emph{superselection rule}. The existence of superselection essentially shows the limitations of quantum coherence, as captured by the interference effect, in a quantum mechanical system.

The first superselection rule discovered by Wick, Wightman and Wigner was in the context of relativistic quantum mechanics exploiting the existence of discrete symmetry of parity.\cite{wick1952} Eventually several superselection rules were discovered, like Bargmann superselection rule in the context of systems with Galilean invariance,\cite{bargmann1954} fermion superselection rule,\cite{wigner1968} and the charge superselection rule. \cite{aharonov1967,wick1970superselection} These superselection rules are of substantial significance owing to their generality. Several other superselection rules have also been found and studied in various context,\cite{giulini2016superselection,wightman1995superselection} like the one found in the one-dimensional hydrogen atom.\cite{yepez1988} The possibility of the existence of superselection due to environmental interactions has also attracted some attention.\cite{zurek1982,giulini1995} In fact the occurrence of chirality in molecular systems in chemistry has been studied from the viewpoint of environmentally induced superselection.\cite{amann1991chirality} The occurrence of superselection and its impact on quantum computation and information aspects has also been explored for a while.\cite{bartlett2007,kitaev2004}

In this paper, we show the existence of a superselection rule, christened \emph{Bloch superselection rule}, in non-relativistic periodic potential problem. This superselection rule dictates that no two Bloch states with different wavevectors can be coherently superposed to yield interference, which captures the relative phase between the two states. It is found that owing to the discrete lattice translation symmetry of the system, the Hilbert space of the system decomposes into disjoint superselection sectors. Each of these superselection sectors is spanned by Bloch states with identical wavevectors albeit from different bands. A topological reason for the existence of Bloch superselection rule is found, and as a result it is believed that this superselection will be robust. This notion of Bloch superselection is also extended to the temporal domain in the context of periodically driven quantum systems, which consists of a wide variety of quantum systems, having the discrete translation symmetry in time. It is found that the Hilbert space of these systems decomposes into superselection sectors spanned by quasi-periodic Floquet states with different quasienergies. As a result these systems do not allow for a coherent superposition of any two quasi-periodic Floquet states with different quasienergies.

In the next section, a brief overview of interference phenomenon in quantum mechanics is presented and the notion of superselection is introduced with an example. In the subsequent section, the Bloch superselection rule discussed, first in the case of periodic lattice systems and later in the context of periodically driven quantum systems. The paper concludes with a brief summary and a discussion of the results.

\section{Interference and lack thereof}

It is a well known fact that, given two generic quantum states $\ket{\j_a}$ and $\ket{\j_b}$, the linear superposition $\ket{\J} = \ket{\j_a} + \ket{\j_b}$ between the two gives rise to the interference effect, which manifests itself in the probability:
\begin{align} \nonumber
\amp{\J}{\J} & =   \amp{\j_a}{\j_a} + \amp{\j_b}{\j_b} \\&+ 2 \: | \amp{\j_a}{\j_b} | \cos (\text{Arg} \: \amp{\j_a}{\j_b}). 
\end{align}
In the literature the term $\text{Arg} \: \amp{\j_a}{\j_b}$ is often referred to as the \emph{relative phase} between the two states $\ket{\j_{a(b)}}$, and is the key term giving rise to periodic extrema in the probability $\amp{\J}{\J}$. \cite{feynmanbook,diracbook,agarwal2013} Evidently this term captures the phase difference between the two states $\ket{\j_{a(b)}}$, since it changes by amount $(\lambda_a - \lambda_b)$, under the phase redefinition $\ket{\j_a} \rightarrow e^{i \lambda_a} \ket{\j_a}$, $\ket{\j_b} \rightarrow e^{i \lambda_b} \ket{\j_b}$, for any real values of $\lambda_{a(b)}$.  When the linearly superposed state $\ket{\J}$ is found to capture the relative phase of the two initial states $\ket{\j_{a(b)}}$ leading to the interference, it is said that a \emph{coherent superposition} of the two states has occurred.\cite{giulini2016superselection, feynmanbook} It is well known that such a notion of coherent superposition does not have any classical counterpart.\cite{diracbook,feynmanbook} 

Note that the interference phenomenon between any two given states does not always manifest in the probability. It is easy to see that when the two states $\ket{\j_{a(b)}}$ are orthogonal, the overlap $\amp{\j_a}{\j_b}=0$, and hence no interference is found in $\amp{\J}{\J}$.  However the interference effect can still manifest itself in the average of some valid observable $\hat{O}(\hat{x},\hat{p})$,  which is also an experimentally measurable quantity. \cite{Note1}
Such an average in the superposed state $\ket{\J} = \ket{\j_a} + \ket{\j_b}$ can be expressed as:
\begin{align} \nonumber
\bra{\J} \hat{O} \ket{\J} = &  \frac{1}{\amp{\J}{\J}} \left( \bra{\j_a} \hat{O} \ket{\j_a} + \bra{\j_b} \hat{O} \ket{\j_b} \right. \\
& \left. + \; 2 \; |\bra{\j_a} \hat{O} \ket{\j_b}| \: \cos (\text{Arg} \bra{\j_a} \hat{O} \ket{\j_b} ) \right).
\end{align}
So long as the magnitude of the overlap $|\bra{\j_a} \hat{O} \ket{\j_b}| \neq 0$, the interference effect is manifested in this average owing to the term $\text{Arg} \; \bra{\j_a} \hat{O} \ket{\j_b}$, which captures the relative phase between the two states. This is evident since it changes by an amount
$(\lambda_a - \lambda_b)$, under the phase redefinition $\ket{\j_a} \rightarrow e^{i \lambda_a} \ket{\j_a}$, $\ket{\j_b} \rightarrow e^{i \lambda_b} \ket{\j_b}$.

If there exists a scenario wherein this overlap term is zero for all the physical observables $\hat{O}$ and identity, then the superposed state $\ket{\J}$ is said to be an \emph{incoherent superposition} or \emph{classical mixture} of states $\ket{\j_a}$ and $\ket{\j_b}$, showing no interference whatsoever in any measurement. \cite{wightman1995superselection,giulini2016superselection} Such mixtures are accurately depicted by the density matrix $\ket{\j_a}\bra{\j_a} + \ket{\j_b} \bra{\j_b}$, which is insensitive to the relative phase between the two states $\ket{\j_{a(b)}}$. This is because the density matrix $\rho = \ket{\J} \bra{\J}$ can be written as $\rho =(\ket{\j_a}\bra{\j_a} + \ket{\j_b} \bra{\j_b}) + (\ket{\j_a} \bra{\j_b} + \ket{\j_b} \bra{\j_a})$. Evidently the relative phase effect is captured by the term $(\ket{\j_a} \bra{\j_b} + \ket{\j_b} \bra{\j_a})$, which gives rise to the overlap $\bra{\j_a} \hat{O} \ket{\j_b}$ in the average $\lan \hat{O} \ran_{\J} =  \text{Tr} \: (\rho \hat{O})/\text{Tr} \; \rho$. Thus in the event when the overlap  $\bra{\j_a} \hat{O} \ket{\j_b} = 0$ for all observables and identity, effectively means that $(\ket{\j_a} \bra{\j_b} + \ket{\j_b} \bra{\j_a})$ is absent from $\rho$. This shows that the state $\ket{\J}$ in effect is identical to the classical mixture described by $\rho = \ket{\j_a}\bra{\j_a} + \ket{\j_b} \bra{\j_b}$. \cite{ecgbook,diracbook,wightman1995superselection}

In quantum physics one also encounters \emph{absolute phase}, which does not have the same stature as the relative phase. The difference between these two phases can be well understood by considering a generic single particle quantum mechanical system. Therein assuming that we are given an orthonormal and complete basis set, denoted as $\{ \ket{a_0}, \ket{a_1}, \cdots \}$, which spans the Hilbert space. A new basis set can be obtained by the action of a unitary operator $\hat{U}_{\lambda}$ on the given basis. Let the unitary operator be defined such that:
\begin{align}  \nonumber
\ket{a'_{n}} &= \hat{U}_{\lambda} \ket{a_n},\\ &= e^{i \lambda} \ket{a_n}, \: \: \: \text{for values of} \: n. \nonumber
\end{align} 
Under such an operation, any state $\ket{v}$ of the Hilbert space, transforms as: 
\begin{align} \label{u2}
\ket{v'} = \hat{U}_{\lambda} \ket{v} = e^{i \lambda} \ket{v}.
\end{align}
The position and momentum operators can be expressed in terms their respective eigenvalues and eigenstates as:
\begin{align} \label{eigenexp}
\hat{x} = \int dx\; x \ket{x} \bra{x}, \: \: \: \text{and} \: \: \hat{p} = \int dp\; p \ket{p} \bra{p},
\end{align}
which shows that under the action of $\hat{U}$ they are invariant:
\begin{align} 
\hat{x} = \hat{U}^{\dagger} \hat{x} \hat{U}\: \: \: \text{and} \: \: \: \hat{p} = \hat{U}^{\dagger} \hat{p} \hat{U}.
\end{align} 
This straightforwardly shows that all the observables are also invariant under this unitary operation, since any general observable $\hat{O}$ is expressible as a function of dynamical variables and time $\hat{O}  =\hat{O}(\hat{x},\hat{p},t)$. This shows that all the observable quantities, like transition amplitude $\amp{v}{w}$ between any two states $\ket{v}$ and $\ket{w}$, and average of any generic observable $\hat{O}$: $\bra{u} \hat{O} \ket{u}/\amp{u}{u}$ in any state $\ket{u}$, remain invariant under the operation $\hat{U}_\lambda$. As a result experimentally is not possible to differentiate whether the system is in the state $\ket{v}$ or $\hat{U}_\lambda \ket{v} \equiv e^{i \lambda} \ket{v}$, for any value of $\lambda$. The existence of this inconsequential albeit inherent phase ambiguity, is referred to as the \emph{absolute phase}.\cite{diracbook,ecgbook}

Consider a non-relativistic quantum system, say an infinite well system, albeit with a possibility of variable number of particles. In such a system, consider two distinct physical states. First wherein a single particle is trapped in the well in some energy level, and let it be denoted as $\ket{A}$. The second case wherein there are two identical particles (assumed to be fermions) trapped in the well in some energy levels, denoted by the state $\ket{BC} = \frac{1}{\sqrt{2}} (\ket{B}_{1} \ket{C}_{2} - \ket{C}_{1} \ket{B}_{2})$. Given these two possible physical states, one wonders about the nature of the possible interference when they are linearly superposed $\ket{A} + \ket{BC}$. In order to probe the interference effect, we consider the average of any general observable $\hat{Q}$ :
\begin{align} \nonumber
&(\bra{A} + \bra{BC}) \hat{Q} (\ket{A} + \ket{BC}) \\ &= \bra{A}\hat{Q}\ket{A} + \bra{BC}\hat{Q}\ket{BC} + 2 \; \text{Re}
\; \bra{A} \hat{Q} \ket{BC}. 
\end{align}
Under the unitary operation $\hat{U}_\lambda$, the states transform as $\ket{A} \ra e^{i \lambda}\ket{A}$
and $\ket{BC} \ra e^{i 2 \lambda}\ket{BC}$. The observable $\hat{Q}$ in general must be a function of two particle dynamical operators $\hat{x}_{1,2}$ and $\hat{p}_{1,2}$  whose action is well defined on two particle state $\ket{BC}$,\cite{Note2} and that of single particle dynamical operators $\hat{X}$ and $\hat{P}$ whose action is well defined on single particle state $\bra{A}$.\cite{diracbook,ecgbook} Expressing these single particle and two particle dynamical operators, as in (\ref{eigenexp}), in terms of their eigenvectors, one immediately sees that they are invariant under the action of $\hat{U}_\lambda$. This implies that all the observables constructed out of these dynamical operators are also invariant. So by resolving the identity, the overlap term can be written as:
\begin{align} \label{overlap1}
\bra{A} \hat{Q} \ket{BC} = \bra{A} \hat{U}^{\dagger} \hat{Q} \hat{U} \ket{BC} = e^{i \lambda} \bra{A} \hat{Q} \ket{BC}.
\end{align}
One could also resolve the identity twice so that: 
\begin{align} \label{overlap2}
\bra{A} \hat{Q} \ket{BC} = \bra{A} \hat{U}^{\dagger} \hat{U}^{\dagger} \hat{Q} \hat{U} \hat{U} \ket{BC} = e^{i 2 \lambda} \bra{A} \hat{Q} \ket{BC}.	\end{align}
Compatibility of (\ref{overlap1}) and (\ref{overlap2}) leads us to the conclusion that the overlap must vanish $\bra{A} \hat{Q} \ket{BC} =0$, in general for all observables. 

For the sake of argument one can point out that there exists a non-trivial self adjoint operator $\bar{O} = (\ket{BC} \bra{A} + \ket{A} \bra{BC})$ such that the overlap $\bra{A} \bar{O} \ket{BC} \neq 0$. This raises the question if the above result is inconsistent or incorrect. A careful reflection will convince the reader that the self adjoint operator $\bar{O}$ can not be a valid observable and should not be considered, since it can not be expressed in terms of dynamical operators like $\hat{X}$, $\hat{P}$ and others. \cite{Note3} This shows that the set of observables corresponding to the system is contained in the set of self adjoint operators defined over the Hilbert space of the system, and not all self adjoint operators qualify to be observables.\cite{giulini2016superselection,wightman1995superselection} 

This discussion shows that the state $\ket{A} + \ket{BC}$ is actually a classical mixture of $\ket{A}$ and $\ket{BC}$, expressible by the density matrix $\ket{A} \bra{A} + \ket{BC} \bra{BC}$, and not a coherent superposition of the two states.

It is easy to see that the same conclusion can be extended to any two states having different number of particles. One finds that while it is possible to coherently superpose two states having same number of particles, it is not possible to do so when the particle number is different. This observation shows that if $\mathscr{V}$ stands for the Hilbert space of the system, then it can be decomposed into a direct sum of subspaces $\mathscr{V}_j$:
\begin{align}
\mathscr{V} = \mathscr{V}_0 \oplus \mathscr{V}_1 \oplus \mathscr{V}_2 \oplus \cdots,
\end{align}
each consisting of all the states having $j$ particles, and their linear combinations. Above discussion shows that any two states both contained within a  subspace $\mathscr{V}_j$ only can be coherently superposed, to give rise to an interference phenomenon. Any superposition of states belonging to different coherent subspaces results only in a classical mixture, devoid of any interference effect. This restriction of interference phenomena is referred to in the literature as \emph{superselection rule}, and the subspaces $\mathscr{V}_j$ are known as \emph{superselection sectors}.\cite{giulini2016superselection} 

The notion of superselection was discovered long back in 1952 by Wick, Wightman and Wigner,\cite{wick1952} has since then generated lot of interest and discussion about its mathematical and physical foundations. The existence of superselection in a system puts a restriction on the quantum behaviour of the system, in the sense that any two states from different superselection sectors can only be added to give an incoherent classical mixture. This fact has generated significant attention and as a result the notion of superselection continues to be carefully studied.\cite{giulini2016superselection,kitaev2004,bartlett2007} 

Over a period of time several superselection rules have been discovered in various contexts,\cite{wick1970superselection,yepez1988,zurek1982} what we have discussed above is an example of fermion superselection rule.\cite{wigner1968,wightman1995superselection}  

\section{Superselection between Bloch waves}

Consider the quantum dynamics of a particle of mass $\mu$ under the influence of a periodic potential.  The Hamiltonian $\hat{H}$ depicting the dynamics of such a system reads:
\begin{align}
\hat{H} = \frac{\hat{p}^2}{2 \mu} + V(\hat{x}),
\end{align}
where $V(\hat{x})$ represents the periodic lattice potential with lattice constant $a$, so that: $V(\hat{x}+a) = V(\hat{x})$. We assume that this system is defined with \emph{periodic boundary condition}(PBC), and the lattice consists of $N$ unit cells. This essentially means that the points $x=0$ and $x=L(=Na)$ are identified. \cite{Note4} Since unit cell is physically identical to rest of the cells, the system possesses discrete translation symmetry under operation $\hat{x} \rightarrow \hat{x} + a$. \cite{Note5}  This symmetry is expressible in terms of the unitary spatial translation operator $\hat{T} = e^{i\frac{\hat{p}}{\hbar}a}$ (where $\hat{T}^{\dagger} \hat{x} \hat{T} = \hat{x} - a$) as:
\begin{align} \label{rel1}
&\hat{T}^{\dagger} \hat{H} \hat{T} = \hat{H}, \: \text{and} \\ \label{rel2} &\hat{T}^{\dagger} \hat{p} \hat{T} = \hat{p}.
\end{align} 
The fact that $[\hat{H}, \hat{T}]=0$ allows for a simultaneous diagonalisation of both these operators, and the wavefunctions $\psi_{n k_{l}}(x) (= \langle x |\psi_{n k_{l}} \rangle$) so obtained are the well known \emph{Bloch waves}.\cite{ashcroftbook,kittelbook} By construction these states solve:
\begin{align} \label{eig1}
&\hat{H} |\psi_{n k_{l}} \rangle = E_{n k_{l}} |\psi_{n k_{l}} \rangle, \: \text{and} \\ \label{eig2}
&\hat{T} |\psi_{n k_{l}} \rangle = e^{i k_{l} a } |\psi_{n k_{l}} \rangle. 
\end{align}    
It is well known that, the spectrum $E_{n k_{l}}$ of the system consists of bands, characterised by the index $n$ ($=1,2,\cdots$), with each band comprising of $(N-1)$ states characterised by wavevector $k_{l} = \frac{2 \pi}{Na}l$ (where $l=0,1,\cdots,N-1$). These Bloch states form an orthonormal and complete basis set over the Hilbert space $\mathscr{H}$ of the system.

Noting the Bloch property of the wavefunction $\psi_{n k_{l}}(x + a) = \psi_{n k_{l}}(x) e^{i k_{l} x}$, it is easy to see that it can be decomposed as: \cite{ashcroftbook,kittelbook}
\begin{align}
\psi_{n k_{l}}(x) = e^{i k_{l} x} u_{n k_{l}}(x), 
\end{align}
where the state $u_{n k_{l}}(x)$ is the \emph{cell periodic Bloch state}, displaying cell periodicity $u_{n k_{l}}(x+a) = u_{n k_{l}}(x)$. This shows that although the domain of the wavefunction $\psi_{n k_{l}}(x) = \langle x |\psi_{n k_{l}} \rangle$ is $[0,Na]$, the nontrivial physical  contribution to it originates from the cell periodic part, captured by $u_{n k_{l}}(x) = \langle x| u_{n k_{l}} \rangle$, whose domain is $[0,a]$. Evidently the states $| u_{n k_{l}} \rangle$ solve the eigenvalue problem for a class of Hamiltonians $\hat{H}_{k_{l}} | u_{n k_{l}} \rangle = E_{n k_{l}} | u_{n k_{l}} \rangle$, where $\hat{H}_{k_{l}} = \frac{(\hat{p} + \hbar k_{l})^2}{2 \mu} + V(\hat{x})$. Thus we see that the cell periodic Bloch states $| u_{n k_{l}} \rangle$ belong to a vector space which is different than the one spanned by $|\psi_{n k_{l}} \rangle$, since the domain and the boundary conditions obeyed by the two wavefunctions are distinct.  Denoting the momentum translation operator as $\hat{T}_{p}(k_l) = e^{i k_{l} \hat{x}}$, the action of any general observable $\hat{\mathscr{O}}(\hat{x},\hat{p})$ on the Bloch state is given by:
\begin{align} \nonumber
\hat{\mathscr{O}}(\hat{x},\hat{p}) \ket{\j_{n k_{l}}} &= \hat{\mathscr{O}}(\hat{x},\hat{p}) \hat{T}_{p}(k_l) \ket{u_{n k_{l}}} \\ \nonumber
& = \hat{T}_{p}(k_l) \hat{\mathscr{O}}(\hat{x},\hat{p} + \hbar k_{l}) \ket{u_{n k_{l}}}.
\end{align}
Now inorder that the action of  $\hat{\mathscr{O}}(\hat{x},\hat{p})$ on the Bloch state in the LHS is well defined, the action of  $\hat{\mathscr{O}}(\hat{x},\hat{p} + \hbar k_l)$ on the $\ket{u_{n k_{l}}}$ needs to be well defined. Now the states $\ket{u_{n k_{l}}}$ obey cell periodicity, and it is required that the operator $\hat{\mathscr{O}}(\hat{x},\hat{p} + \hbar k_{l})$ must also obey the cell periodicity condition 
\begin{align} \label{obs}
\hat{\mathscr{O}}(\hat{x},\hat{p}) = \hat{\mathscr{O}}(\hat{x}-a,\hat{p}) \equiv \hat{T} \hat{\mathscr{O}}(\hat{x},\hat{p}) \hat{T}^{\dagger},
\end{align}
in order to be a valid linear operator acting on $\ket{u_{n k_{l}}}$.  This necessary condition ensures that the state $\hat{\mathscr{O}}(\hat{x},\hat{p} + \hbar k_{l}) \ket{u_{n k_{l}}}$ and $\ket{u_{n k_{l}}}$ belong to the same vector space.

Now let us consider a superposition $| \Phi \rangle$ of any two generic Bloch states $|\psi_{m k_{j}}\rangle$ and $|\psi_{n k_{l}}\rangle$:
\begin{align}
| \Phi \rangle = |\psi_{m k_{j}}\rangle +  |\psi_{n k_{l}}\rangle. 
\end{align} 
The average of any generic observable $\hat{\mathscr{O}}(\hat{x},\hat{p})$ in the state $|\Phi \rangle$ is given by:
\begin{align} \nonumber
{\langle \Phi | \hat{\mathscr{O}} | \Phi \rangle} & =  \frac{1}{\amp{\Phi}{\Phi}}  \left( \langle \psi_{m k_{j}}| \hat{\mathscr{O}} |\psi_{m k_{j}}\rangle + \langle \psi_{n k_{l}}| \hat{\mathscr{O}} |\psi_{n k_{l}}\rangle \right) \\
& \:+  \left( \langle \psi_{m k_{j}}| \hat{\mathscr{O}} |\psi_{n k_{l}}\rangle + \text{complex conjugate} \right).
\end{align}
Let us look at the interference term $\langle \psi_{m k_{j}}| \hat{\mathscr{O}} |\psi_{n k_{l}}\rangle$ carefully. Using the spatial translation operator $\hat{T}$ this can be rewritten as:
\begin{align}
\langle \psi_{m k_{j}}| \hat{\mathscr{O}} |\psi_{n k_{l}}\rangle = \langle \psi_{m k_{j}}|\hat{T}^{\dagger} \left( \hat{T}\hat{\mathscr{O}} \hat{T}^{\dagger} \right) \hat{T} |\psi_{n k_{l}}\rangle.
\end{align}
Using (\ref{eig2}) and (\ref{obs}) in RHS leads us to a very important relation:
\begin{align}
\langle \psi_{m k_{j}}| \hat{\mathscr{O}} |\psi_{n k_{l}}\rangle = e^{i (k_{l} - k_{j}) a } \times \langle \psi_{m k_{j}}| \hat{\mathscr{O}} |\psi_{n k_{l}}\rangle. 
\end{align} 
So for all cases when $k_j \neq k_l$, this relation implies that the interference term $\langle \psi_{m k_{j}}| \hat{\mathscr{O}} |\psi_{n k_{l}}\rangle = 0$ for all the physical observables.  In the light of the discussions in the previous section, we immediately see that the linear superposition of any two Bloch states with different wave vectors do not give rise to interference, but to rather an incoherent classical mixture described by density matrix $\ket{\Phi} \bra{\Phi} = \ket{\j_{n k_{l}}} \bra{\j_{n k_{l}}} + \ket{\j_{m k_{j}}} \bra{\j_{m k_{j}}}$.

One has thus unveiled \emph{a superselection rule}, which we refer to as \emph{Bloch superselection rule} operating in this system: \emph{Bloch waves with different wave vectors can not be coherently superposed}. The Hilbert space $\mathscr{H}$ of the system decomposes into direct sum of coherent subspaces $\mathscr{H}_{k_{j}}$ as:
\begin{align}
\mathscr{H} = \mathscr{H}_{k_{0}} \oplus \mathscr{H}_{k_{1}} \oplus \dots \oplus \mathscr{H}_{k_{N-1}}.   
\end{align}
Each coherent subspace $\mathscr{H}_{k_{j}}$ is spanned by the basis set $\{|\psi_{0 k_{j}}\rangle, |\psi_{1 k_{j}}\rangle, \cdots \}$ consisting of Bloch states with different band index $n$ but with fixed wavevector $k_{j}$. 

Actually this superselection rule is of topological origin, which can be seen by noting the fact that $\psi_{n k_{l}}(x) = e^{i k_{l} x} u_{n k_{l}}(x)$ and the factor $e^{i k_{l} x}$ is a topologically nontrivial object. It is a function of $x$ respecting the PBC: $e^{i k_{l} x} = e^{i k_{l} (x + L)}$. This shows that such a function lives on a coordinate space which is actually a circle with circumference $L$. The object $e^{i k_{l} x}$ by definition is a complex number with unit modulus, and takes values only on the unit circle in the complex plane. So $e^{i k_{l} x} = e^{i \frac{2 \pi }{L}l x}$ is a map from spatial circle (with circumference $L$) to the unit circle. Mathematically, such maps are characterized by a topological index, an integer called the \emph{winding number}, which measures the number of times one circle is winded on another.\cite{nakahara2003} It is easy to see that the integer $l$ is actually the winding number; under one circuit in space, the factor $e^{i \frac{2 \pi }{L}lx}$ completes $l$ rotations on the unit circle. Evidently it is not possible to continuously deform $e^{i \frac{2 \pi}{L}lx}$ to some $e^{i \frac{2 \pi}{L}jx}$ ($l \neq j$). This shows that the functions having different winding numbers are fundamentally distinct. So while the two Bloch waves with same winding number $l$: $\psi_{m k_{l}}(x)$ and $\psi_{n k_{l}}(x)$ can be coherently added, two Bloch waves with different winding numbers do not admit such an addition. 

Having understood the origin and workings of the Bloch superselection rule one may wonder if this superselection rule can provide some other observationally significant insight into the system at hand. Interestingly this superselection rule indeed readily provides some quick insights when it is employed with sum rules. \cite{wang1999} To demonstrate this point let us consider the sum rule for the average of square of some generic observable $\hat{\mathscr{O}}$ in the Bloch state $|\psi_{n k_{l}}\ran$:
\begin{align}
	\lan \psi_{n k_{l}} | \hat{\mathscr{O}}^2 | \psi_{n k_{l}} \ran = \sum_{m, k_{j}} \lan \psi_{n k_{l}} | \hat{\mathscr{O}} | \psi_{m k_{j}} \ran \lan \psi_{m k_{j}} | \hat{\mathscr{O}} | \psi_{n k_{l}} \ran.
\end{align}
Owing to the superselection one immediately sees that the amplitude $\lan \psi_{m k_{j}} | \hat{\mathscr{O}} | \psi_{n k_{l}} \ran$ vanishes whenever $k_{j} \neq k_{l}$, which allows this expression to be written as:
\begin{align*}
\lan \psi_{n k_{l}} | \hat{\mathscr{O}}^2 | \psi_{n k_{l}} \ran &= \lan \psi_{n k_{l}} | \hat{\mathscr{O}} | \psi_{n k_{l}} \ran \lan \psi_{n k_{l}} | \hat{\mathscr{O}} | \psi_{n k_{l}} \ran	\\ &+ \sum_{m \neq n} \lan \psi_{n k_{l}} | \hat{\mathscr{O}} | \psi_{m k_{l}} \ran \lan \psi_{m k_{l}} | \hat{\mathscr{O}} | \psi_{n k_{l}} \ran.
\end{align*} 
Noticing that the square of uncertainty $\Delta \mathscr{O}$ of the observable $\hat{\mathscr{O}}$ in the state $|\psi_{n k_{l}}\ran$ is given by $\Delta \mathscr{O}_{n k_{l}}^2 = \lan \psi_{n k_{l}} | \hat{\mathscr{O}}^2 | \psi_{n k_{l}} \ran - \lan \psi_{n k_{l}} | \hat{\mathscr{O}} | \psi_{n k_{l}} \ran^{2}$, the above expression reads:
\begin{align}
\Delta \mathscr{O}_{n k_{l}}^2 = \sum_{m \neq n} |\lan \psi_{m k_{l}} | \hat{\mathscr{O}} | \psi_{n k_{l}} \ran|^{2}.
\end{align}  
This is a very interesting sum rule, and let us consider the case when the system has only one band like the single band tight binding model. \cite{ashcroftbook} This sum rule then immediately dictates that $\Delta \mathscr{O}_{n k_{l}}^2 = 0$. When the observable at hand is momentum $\hat{p}$, this immediately tells us that the momentum uncertainty vanishes in the Bloch state $\Delta {p}_{n k_{l}}^2 = 0$. This asserts that the Bloch states in the one band lattice model are actually momentum eigenstates. This is a well known fact, which one discovers only after diagonalising the model. \cite{ashcroftbook} However from the sum rule and superselection rule, one readily obtains this result without any elaborate calculation.

\section{Superselection between Floquet states}

From the discussion presented in the above section, it is clear that it is the existence of discrete translation symmetry of the lattice potential which gives rise to the superselection rule for the Bloch states. It is well known that there exists a temporal analogue of the Bloch states, which are known as Floquet states,\cite{tannor2018,grifoni1998,shirley1965} and below we show that a similar superselection rule also holds for these states as well. 

Consider a general quantum system which is externally driven by a periodic field, so that its Hamiltonian $\hat{H}(t)$ is periodic in time: $\hat{H}(t+T) = \hat{H}(t)$, where $T$ is the period. 
By this assumption we are addressing  a large class of periodically driven systems, ranging from a system with finite dimensional vector space like the one described by: $\hat{H}(t) = \epsilon \sigma_z + \sin (\omega t) \sigma_x$, to the one defined in infinite dimensional space as $\hat{H}(t) = \hat{p}^2 + \hat{x}^2 + \sin (\omega t) \hat{x}$. The time evolution of some generic quantum state $\ket{\Phi(t)}$ is governed by the Schrodinger equation:
\begin{align} \label{scr}
\hat{H}(t) \ket{\Phi(t)} = i \hbar \frac{\partial}{\partial t} \ket{\Phi(t)}. 
\end{align}
Let us consider our attention on a class of states, which we refer to as quasi-periodic Floquet states $\ket{\phi_\epsilon (t)}$, because they obey the Bloch property in time evolution over a period:
\begin{align} \label{qpproperty}
\ket{\phi_\epsilon (t + T)} = \hat{U}(T)\ket{\phi_\epsilon(t)} = e^{- i \epsilon T/\hbar} \ket{\phi_\epsilon(t)},
\end{align} 
where $\hat{U}(T)$ is the unitary time evolution operator over a period.  Note that this relation is the temporal analogue of the Bloch property, with the role of wavevector $k$ being played by $\epsilon$, which is called the quasienergy. Akin to the wavevector definition, the quasienergies are only defined modulo an integer multiple of quantum $\hbar \omega$, where $\omega = 2 \pi/T$. 
It is worth mentioning that the set of quasi-periodic Floquet states $\{ \ket{\phi_\epsilon(t)} \}$ form a basis for the underlying vector space of the system, owing to the fact that these stats solve the eigenvalue problem for a unitary operator. This knowledge can be used to describe the time evolution of a quantum state $\ket{\Phi(t)}$ as a linear superposition of the quasi-periodic Floquet states:
\begin{align} \label{linearc}
	\ket{\Phi(t)} = \sum_{\epsilon} c_{\epsilon} \ket{\phi_\epsilon(t)},
\end{align}  
where the coefficients $c_{\epsilon}$ are complex constants.

It is beneficial to decompose the state $\ket{\phi_\epsilon(t)}$ as:  
\begin{align} \label{floquet}
\ket{\phi_\epsilon(t)} = e^{- i \epsilon t/\hbar} \ket{v_\epsilon(t)},
\end{align}
where the periodic Floquet state $\ket{v_\epsilon(t)}$ is the analogue of cell periodic Bloch state, and obeys the condition: $\ket{v_\epsilon(t + T)} = \ket{v_\epsilon(t)}$.\cite{tannor2018,grifoni1998} Plugging this relation for the Floquet state $\ket{v_\epsilon(t)}$ in the Schrodinger equation gives us an eigenvalue problem for these states:
\begin{align}
\left( \hat{H}(t) - i \hbar \frac{\partial}{\partial t}\right) \ket{v_\epsilon(t)} = \epsilon \ket{v_\epsilon(t)}.
\end{align}
In practice it is this eigenvalue problem that is solved to yield the quasienergy spectrum and the periodic Floquet states. \cite{shirley1965,tannor2018,grifoni1998} 

It is worth noting that owing to the quasi-periodic property (\ref{qpproperty}) of the state $\ket{\phi_{\epsilon}(t)}$, the average $\bra{\phi_\epsilon (t)} \hat{O} \ket{\phi_{\epsilon}(t)}$ of any general observable $\hat{O}(\hat{x},\hat{p})$ can be expressed as:
\begin{align*}
	\bra{\phi_\epsilon (t)} \hat{O} \ket{\phi_{\epsilon}(t)} &= \bra{\phi_\epsilon (t + T)} \hat{O} \ket{\phi_{\epsilon}(t + T)} \\ &\equiv \bra{\phi_\epsilon (t)} \hat{U}^{\dagger}(T) \hat{O} \hat{U}(T) \ket{\phi_{\epsilon}(t)},	
\end{align*}
as a result all the admissible observables in the theory must respect the temporal periodicity:
\begin{align}
	\hat{O} = \hat{U}^{\dagger}(T) \hat{O} \hat{U}(T).
\end{align}

As noted in the earlier section, the effect of interference due to the linear superposition of two quasi-periodic Floquet states $\ket{\phi_\epsilon(t)}$ and $\ket{\phi_{\epsilon'} (t)}$ with different quasienergies, is encoded in the overlap $\langle  \phi_\epsilon (t) | \hat{O} |  \phi_{\epsilon'} (t) \rangle$, for any general observable $\hat{O}(\hat{x},\hat{p})$. It is immediately clear that the overlap $\langle  \phi_\epsilon (t) | \hat{O} |  \phi_{\epsilon'} (t) \rangle = 0$ for all $\epsilon \neq \epsilon'$ since: 
\begin{align} \nonumber
\langle  \phi_\epsilon (t) | \hat{O} |  \phi_{\epsilon'} (t) \rangle &=
\langle  \phi_\epsilon (t) | \hat{U}^{\dagger}(T) \hat{U}(T) \hat{O} \hat{U}^{\dagger}(T) \hat{U}(T) |  \phi_{\epsilon'} (t) \rangle \nonumber \\ &= e^{i (\epsilon' - \epsilon) T/\hbar} \langle  \phi_\epsilon (t) | \hat{O} |  \phi_{\epsilon'} (t) \rangle. \nonumber
\end{align}
Thus we see \emph{the existence of a temporal analogue of the Bloch superselection rule}. This superselection rule implies that \emph{a linear superposition of two quasi-periodic Floquet states with different quasienergies does not lead to their coherent superposition/interference}. Rather such an addition leads to an incoherent classical mixture of the quasi-periodic Floquet states. Thus the density matrix $\ket{\Phi(t)} \bra{\Phi(t)}$ corresponding to the general state $\ket{\Phi(t)}$ (as given by (\ref{linearc})) is expressible as:
\begin{align*}
\ket{\Phi(t)} \bra{\Phi(t)} = \sum_{\epsilon} |c_{\epsilon}|^{2} \ket{\phi_{\epsilon}(t)} \bra{\phi_{\epsilon}(t)}.
\end{align*}
Interestingly one notes that each quasi-periodic Floquet state lives in a superselection sector of its own, in contrast to the earlier encountered Bloch states.

From this discussion it is clear that if the state of the driven system governed by $\hat{H}(t)$ is expressible as a linear superposition of quasi-periodic Floquet states (as given by (\ref{linearc})), then the system is essentially an incoherent classical mixture of the quasi-periodic Floquet states. This important result is bound to have significant implications on the aspects of coherence and tunneling in these driven systems.\cite{deng2015,grifoni1998,dalibard2014}

\section{Conclusion}

In this paper, we have shown the existence of a superselection rule, called the Bloch superselection rule, in the well studied quantum system of a particle in a periodic potential. This superselection rule forbids any kind of interference between two Bloch states with different wavevectors. The existence of this superselection is due to the discrete translation symmetry of the lattice, and it is found to be of topological nature. The fact that this superselection rule is protected by a topological property makes it robust, and one can expect Bloch superselection rule to hold even in the presence of small perturbations, which may be present in the system due to disorder and environmental interactions. It must be pointed out, that although the treatment presented here is for a one dimensional lattice system, the same result would also follow for higher dimensional lattice systems. 

From this superselection rule one immediately learns that the well known Wannier states,\cite{ashcroftbook} which are a particular class of weighted superpositions of all the Bloch states in a band, are actually a classical incoherent mixture of the Bloch states. Similar conclusion can instantly be drawn about other localised wave packets often used to describe ground states in condensed matter systems. It is hoped that these results will throw light on the nature of coherence in quantum condensed matter systems. 

Subsequently we investigate periodically driven quantum systems, and show that there exists a temporal analogue of the Bloch superselection rule in these class of systems. This rule dictates that no two quasi-periodic Floquet states (which are temporal analogues of Bloch states) with different quasi-energies can be coherently superposed, since each state resides in a different superselection sector. The existence of such a superselection puts a severe limitation on the quantum nature in these systems. Interestingly there exists a wide variety of externally driven systems, ranging from two level atom to atomic Bose-Einstein condensates, all coming under the purview of this superselection.\cite{dalibard2014,deng2015,meinert2016,oka2019}

It would be interesting to study the implications of both of these superselection rules simultaneously in the context of periodically driven lattice systems.\cite{demler2010,oka2019} A particularly interesting model to understand the physical manifestation of these rules is the one considered by Thouless,\cite{thouless1983} and lately experimentally realised,\cite{lohse2016} which is known to show quantised particle transport due to topological reasons.

\section*{Acknowledgments}
The author acknowledges beneficial discussions with Prof. J. Samuel and Dr. D. Roy from Raman Research Institute, as also with Prof. P. K. Panigrahi from IISER Kolkata.

\end{document}